\documentclass[12pt,english]{article}
\usepackage[T1]{fontenc}
\usepackage[latin1]{inputenc}
\usepackage{a4wide}
\usepackage{babel}
\usepackage{color}
\usepackage{graphics}

\makeatletter

\providecommand{\LyX}{L\kern-.1667em\lower.25em\hbox{Y}\kern-.125emX\@}

 \newcommand{\lyxaddress}[1]{
   \par {\raggedright #1 
   \vspace{1.4em}
   \noindent\par}
 }

\makeatother
\begin{document}
\selectlanguage{english}

\title{Boundary sine-Gordon model}

\author{\textcolor{black}{\normalsize Z. Bajnok, L. Palla and G. Tak\'acs }}

\maketitle

\lyxaddress{\centering \textcolor{black}{\emph{\small Institute for Theoretical
Physics}}\\
\textcolor{black}{\emph{\small Roland E\"otv\"os University Budapest
}}\\
\textcolor{black}{\emph{\small H-1117 P\'azm\'any s\'et\'any 1/A,
Hungary}}}

\begin{abstract}
We review our recent results on the on-shell description of sine-Gordon
model with integrable boundary conditions. We determined the spectrum
of boundary states together with their reflection factors by closing
the boundary bootstrap and checked these results against WKB quantization
and numerical finite volume spectra obtained from the truncated conformal
space approach. The relation between a boundary resonance state and
the semiclassical instability of a static classical solution is analyzed
in detail. 
\end{abstract}

\section*{Introduction}

Sine-Gordon field theory is defined by the Lagrangean \begin{equation}
\label{bulksgl}
\mathcal{L}=\frac{1}{2}(\partial \Phi )^{2}+\frac{m^{2}}{\beta ^{2}}\cos (\beta \Phi )\quad ,
\end{equation}
where \( \Phi  \) is a real scalar field and \( \beta  \) is the
coupling constant. It is one of the most important quantum field theoretic
models with numerous applications ranging from particle theoretic
problems to condensed matter systems, and one which has played a central
role in our understanding of \( 1+1 \) dimensional field theories.
A crucial property of the model is integrability, which permits an
exact analytic determination of many of its physical properties and
characteristic quantities. 

Integrability can also be maintained in the presence of boundaries
\cite{Skl}; for sine-Gordon theory, the most general boundary potential
that preserves integrability was found by Ghoshal and Zamolodchikov
\cite{GZ}\begin{equation}
\label{bpot}
V_{B}=M_{0}\cos \left( \frac{\beta }{2}(\Phi (0)-\varphi _{0})\right) \quad .
\end{equation}
 They also introduced the notion of 'boundary crossing unitarity',
and combining it with the boundary version of the Yang-Baxter equations
they were able to determine soliton reflection factors on the ground
state boundary. Later Ghoshal completed this work by determining the
breather reflection factors \cite{G} using a boundary bootstrap equation
first proposed by Fring and K\"oberle \cite{FK}.

The first (partial) results on the spectrum of the excited boundary
states were obtained by Saleur and Skorik for Dirichlet boundary conditions
\cite{SkS}. However, they did not take into account the boundary
analogue of the Coleman-Thun mechanism, the importance of which was
first emphasized by Dorey et al. \cite{bct}. Using this mechanism
Mattsson and Dorey were able to close the bootstrap in the Dirichlet
case and determine the complete spectrum and the reflection factors
on the excited boundary states \cite{Dir}. Recently we used their
ideas to obtain the spectrum of excited boundary states and their
reflection factors for the Neumann boundary condition \cite{Neu}
and then for the general two-parameter family of integrable boundary
conditions \cite{Genspec}.

Another interesting problem is the relation between the ultraviolet
(UV) parameters that appear in the perturbed CFT Hamiltonian and the
infrared (IR) parameters in the reflection factors. This relation
was first obtained by Al. B. Zamolodchikov \cite{Aup} together with
a formula for the boundary energy; however, his results remain unpublished.
In order to have these formulae, we rederived them in our paper \cite{uvir},
where we used them to check the consistency of the spectrum and of
the reflection factors against a boundary version of truncated conformal
space approach (TCSA). Combining the TCSA results with analytic methods
of the Bethe Ansatz we found strong evidence that our understanding
of the spectrum of boundary sine-Gordon model is indeed correct.

Recently M. Kormos and one of us (L.P.) achieved the semiclassical
quantization of the two lowest energy static solutions of the model
\cite{KP}. By comparing the quantum corrected energy with the exact
one the perturbative correspondence between the Lagrangean and the
bootstrap parameters has been established. In the paper we extend
their analysis for an unstable solution which corresponds to a boundary
resonance state. We compute the decay rate and decay width of the
resonance state and show how these results agree with the semiclassical
considerations. We comment also on the possible changes in the finite
volume spectra due to the resonance state. 

The paper is organized as follows: in Section 2 we review the boostrap
philosophy by applying to the bulk sine-Gordon theory. In Section
3 we give the boundary analogue of this picture, the boundary spectrum
is determined and the boundary bootstrap is closed by explaining any
pole in the reflection matrix either as a new boundary state or by
the boundary analogue of the Coleman-Thun mechanism. In Section 4
we check the boundary spectrum and reflection factors against finite
volume spectra. Finally in Section 5 we analyze the semiclassical
issues and conclude in Section 6.

\section*{Bootstrap in the bulk sine-Gordon theory}

The bulk sine-Gordon theory described by (\ref{bulksgl}) is an integrable
model since it has infinitely many conserved quantities. As a consequence,
there is no particle production in the scatterings and the multiparticle
S-matrix factorizes into the product of two particle S-matrices, which
can be computed using the requirements of unitarity, crossing symmetry
and the Yang-Baxter equation modulo CDD type ambiguity. The most general
solution for the scattering of an \( O(2) \) symmetric doublet has
the following form\textcolor{black}{\begin{eqnarray*}
S(\lambda ,\theta ) & = & \left( \begin{array}{cccc}
-1 & 0 & 0 & 0\\
0 & \frac{-\sin \lambda \pi }{\sin \lambda (\pi +i\theta )} & \frac{\sin i\lambda \theta }{\sin \lambda (\pi +i\theta )} & 0\\
0 & \frac{\sin i\lambda \theta }{\sin \lambda (\pi +i\theta )} & \frac{-\sin \lambda \pi }{\sin \lambda (\pi +i\theta )} & 0\\
0 & 0 & 0 & -1
\end{array}\right) \times \\
 &  & \prod ^{\infty }_{l=1}\left[ \frac{\Gamma (2(l-1)\lambda +\frac{\lambda i\theta }{\pi })\Gamma (2l\lambda +1+\frac{\lambda i\theta }{\pi })}{\Gamma ((2l-1)\lambda +\frac{\lambda i\theta }{\pi })\Gamma ((2l-1)\lambda +1+\frac{\lambda i\theta }{\pi })}/(\theta \to -\theta )\right] \quad ,
\end{eqnarray*}
where \( \lambda  \) is a free parameter. It describes the scattering
of the soliton anti-soliton doublet \( (s,\bar{s}) \). }

\textcolor{black}{The poles of the scattering matrix signal the existence
of other particles appearing as bound states. For the soliton anti-soliton
scattering they are located at \( \theta =i(\pi /2-u_{n})=i\pi /2-\frac{in\pi }{2\lambda } \).
The corresponding particles are called breathers \( B_{n} \) and
have masses \( m_{B_{n}}=2M\sin (u_{n}) \). }

\textcolor{black}{The scattering matrix of the breathers on the soliton
doublet can be computed from the bootstrap procedure, which graphically
looks as follows:\vspace{1cm}}

\resizebox*{15cm}{!}{\includegraphics{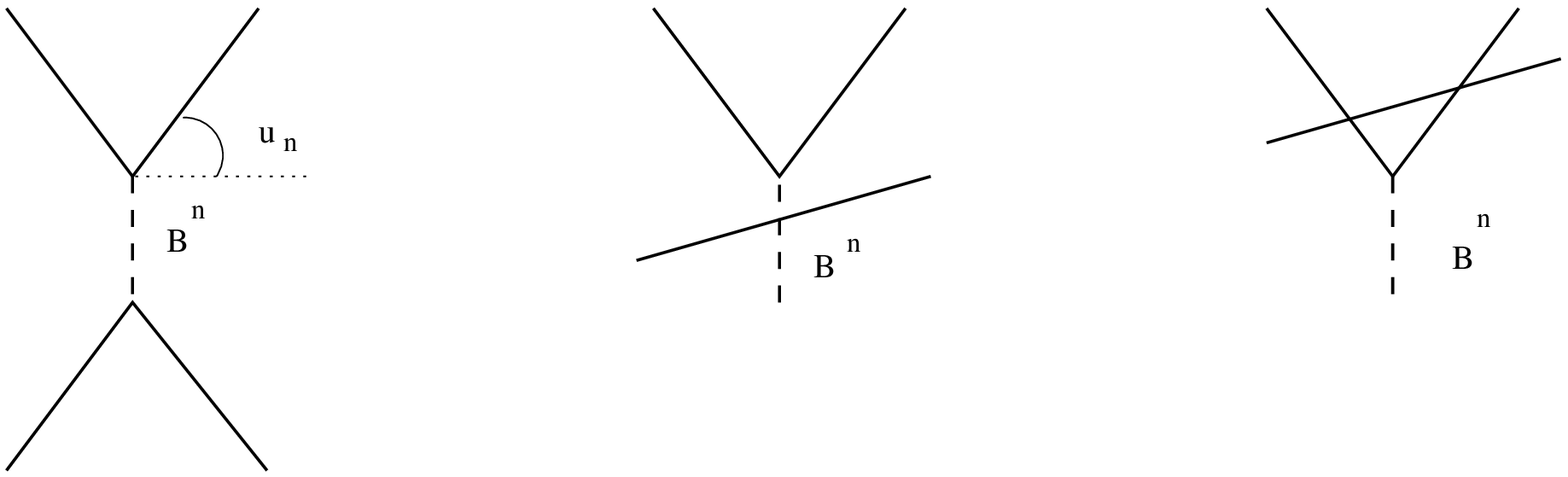}} 

\vspace{.5cm}

Soliton (anti-soliton) lines are shown as straight lines, while the
dashed ones correspond to breathers. The result turns out to be \[
S^{n}=\{n-1+\lambda \}\{n-3+\lambda \}\dots \quad \left\{ \begin{array}{c}
\{1+\lambda \}\quad \textrm{if}\, \, n\: \textrm{is even}\\
-\sqrt{\{\lambda \}}\quad \textrm{if}\, \, n\: \textrm{is odd}
\end{array}\right. \quad ,\]
where \[
\{y\}=\frac{\left( \frac{y+1}{2\lambda }\right) \left( \frac{y-1}{2\lambda }\right) }{\left( \frac{y+1}{2\lambda }-1\right) \left( \frac{y-1}{2\lambda }+1\right) }\quad ;\quad (x)=\frac{\sin (i\theta /2-x\pi /2)}{\sin (i\theta /2+x\pi /2)}\quad .\]
Analysing the pole structure of the breather-soliton/anti-soliton
scattering matrix we find poles, which can be explained by soliton/anti-soliton
intermediate states.

The breather-breather scattering matrix can be computed again by applying
the fusion procedure, but now for the breather-soliton scattering
matrix. The result has the following compact form:\textcolor{black}{\begin{equation}
\label{snm}
S^{n,m}=\{n+m-1\}\{n+m-3\}\dots \{n-m+3\}\{n-m+1\}\quad n\geq m\quad .
\end{equation}
The poles of the scattering matrix (\ref{snm}) can be explained either
by breather intermediate states or by Coleman-Thun type mechanism
as illustrated on the following figure:\vspace{1cm}}

~~~~~~~~~\resizebox*{!}{5cm}{\includegraphics{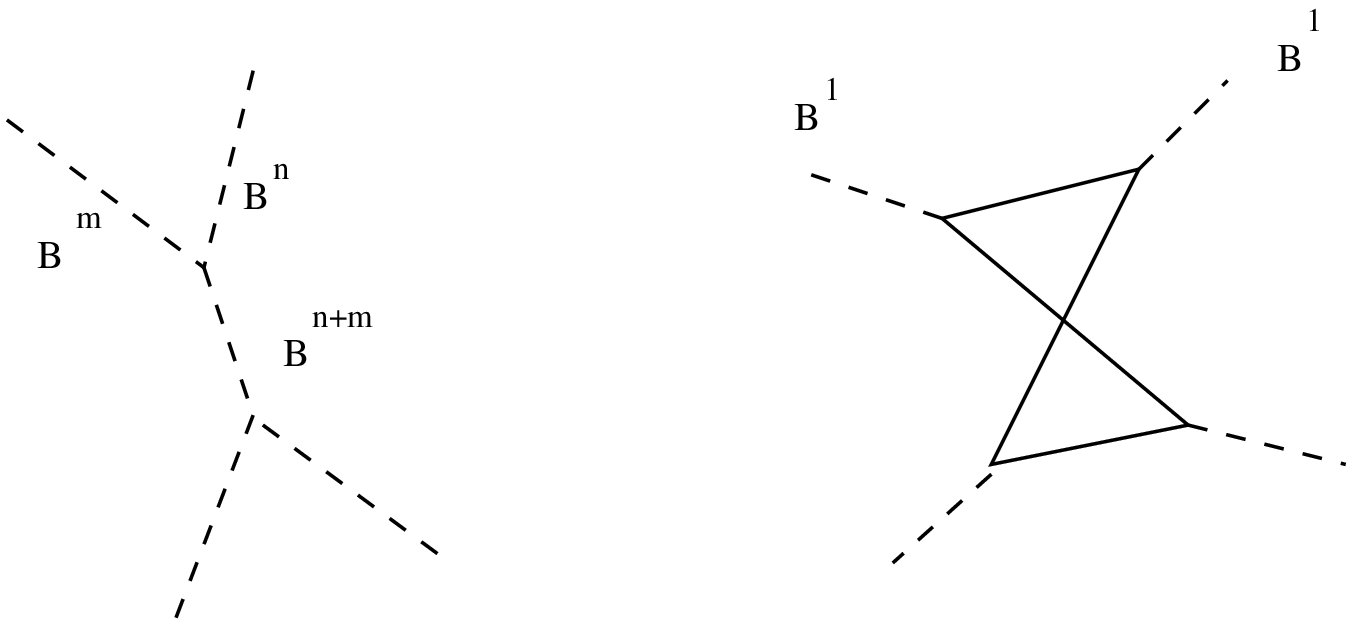}} 

\textcolor{black}{\vspace{.5cm}}

Since we explained all the poles of all the scattering matrices the
sine-Gordon model is solved in the bootstrap sense. This solution
has, however, no clear relation to the Lagrangean. In order to relate
the bootstrap parameters to the parameters of the Lagrangean an alternative
analysis is needed. Performing the comparison of Thermodynamic Bethe
ansatz with conformal perturbation theory \cite{Zmg} or carrying
out the semiclassical quantization of the model \cite{DHN} the exact
mapping can be obtained \[
\lambda =\frac{8\pi }{\beta ^{2}}-1\quad ;\quad M=m^{\frac{8\pi }{8\pi -\beta ^{2}}}\kappa (\beta )\quad ,\]
where the actual form of \( \kappa (\beta ) \) depends on the scheme
in which the quantum theory is defined. In the conformal perturbation
framework \[
\kappa (\beta )=\frac{2\Gamma \left( \frac{\beta ^{2}}{2(8\pi -\beta ^{2})}\right) }{\sqrt{\pi }\Gamma \left( \frac{4\pi }{8\pi -\beta ^{2}}\right) }\left[ \frac{\pi \Gamma \left( 1-\frac{\beta ^{2}}{2\pi }\right) }{2\beta ^{2}\Gamma \left( \frac{\beta ^{2}}{8\pi }\right) }\right] ^{\frac{4\pi }{8\pi -\beta ^{2}}}\quad .\]

\section*{Bootstrap in the boundary sine-Gordon theory}

The boundary sine-Gordon theory can be obtained by restricting the
bulk sine-Gordon theory, (\ref{bulksgl}), to the negative half line
and imposing the most general integrability preserving boundary condition
at the origin \[
\partial _{x}\Phi (x,t)|_{x=0}=-\frac{dV_{B}(\Phi )}{d\Phi }\quad .\]
Due to the integrability of the model the allowed physical processes
are highly restricted. Besides the usual bulk constraints we also
have factorized and elastic reflection on the boundary. Moreover,
the one particle reflection matrices have to obey unitarity, boundary
Yang-Baxter equations and boundary crossing. The most general solution
contains two parameters, similarly to the boundary potential, and
has the following form\textcolor{black}{\[
R(\lambda ,\eta ,\Theta )=\left( \begin{array}{cc}
P^{+} & Q\\
Q & P^{-}
\end{array}\right) R_{0}(\theta )\frac{\sigma (\eta ,\theta )}{\cos \eta }\frac{\sigma (i\Theta ,\theta )}{\cosh \Theta }\]
\[
P^{\pm }=\cos (i\lambda \theta )\cos \eta \cosh \Theta \pm (\cos \leftrightarrow \sin )\quad ;\quad Q=\cos i\lambda \theta \sin i\lambda \theta \quad ,\]
 where \[
R_{0}(\theta )=\prod _{l=1}^{\infty }\left[ \frac{\Gamma (4l\lambda +2i\lambda \theta /\pi )\Gamma (4(l-1)\lambda +1+2i\lambda \theta /\pi )}{\Gamma ((4l-3)\lambda +2i\lambda \theta /\pi )\Gamma ((4l-1)\lambda +1+2i\lambda \theta /\pi )}/(\theta \to -\theta )\right] \]
 is the boundary condition independent part and \[
\sigma (x,\theta )=\frac{\cos (x)}{\cos (x-i\lambda \theta )}\prod _{l=1}^{\infty }\left[ \frac{\Gamma (\frac{1}{2}+\frac{x}{\pi }+(2l-1)\lambda +\frac{i\lambda \theta }{\pi })\Gamma (\frac{1}{2}-\frac{x}{\pi }+(2l-1)\lambda +\frac{i\lambda \theta }{\pi })}{\Gamma (\frac{1}{2}-\frac{x}{\pi }+(2l-2)\lambda +\frac{i\lambda \theta }{\pi })\Gamma (\frac{1}{2}+\frac{x}{\pi }+2l\lambda +\frac{i\lambda \theta }{\pi })}/(\theta \to -\theta )\right] \]
describes the boundary condition dependence. }

\textcolor{black}{The poles of these amplitudes signal the presence
of boundary bound states. The boundary independent poles of \( R_{0}(\theta ) \)
have explanations in terms of boundary Coleman-Thun mechanism \cite{Genspec}.
The boundary dependent poles at \[
\theta =i\nu _{n}=i\left( \frac{\eta }{\lambda }-\frac{(2n+1)}{2\lambda }\right) \]
correspond to bound states, denoted by \( |n\rangle  \) with energy
\( m_{|n\rangle }=M\cos (\nu _{n}) \). The reflection factors on
these boundary bound states can be computed from the bootstrap principle,
which graphically looks as follows:}\textcolor{black}{\small \vspace{.5cm}}{\small \par}

\vspace{0.3cm}
{\noindent \centering \resizebox*{15cm}{!}{\includegraphics{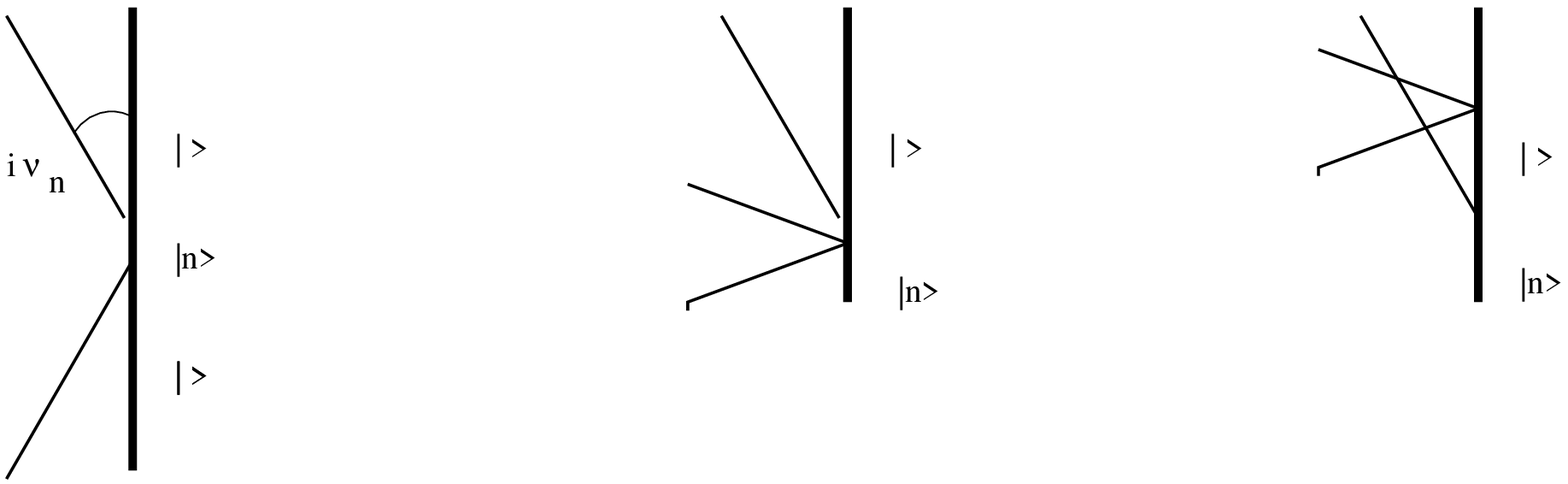}} \par}
\vspace{0.3cm}

\textcolor{black}{\vspace{.5cm}}

The result of the computation is \textcolor{black}{\[
R_{|n\rangle }(\lambda ,\eta ,\Theta )=\bar{R}(\lambda ,\bar{\eta },\Theta )a_{n}(\eta ,\theta )\quad ;\quad a_{n}(\eta ,\theta )=\prod _{l=1}^{n}\left\{ 2\left( \frac{\eta }{\pi }-l\right) \right\} \quad ,\]
where in \( \bar{R} \) the solitons and the anti-solitons are changed
as \( \bar{P}^{\pm }(\lambda ,\eta ,\Theta )=P^{\mp }(\lambda ,\bar{\eta },\Theta ) \),
\( Q(\lambda ,\eta ,\Theta )=Q(\lambda ,\bar{\eta },\Theta ) \) and
\( \bar{\eta }=\pi (\lambda +1)-\eta  \). Analyzing the pole structure
of these excited reflection factors we find poles at \( \theta =iw_{m}=i\nu _{m}(\bar{\eta }) \)
and at \( \theta =i\nu _{n-k} \), with the following boundary Coleman-Thun
explanations:}

\vspace{0.3cm}
{\centering \resizebox*{!}{5cm}{\includegraphics{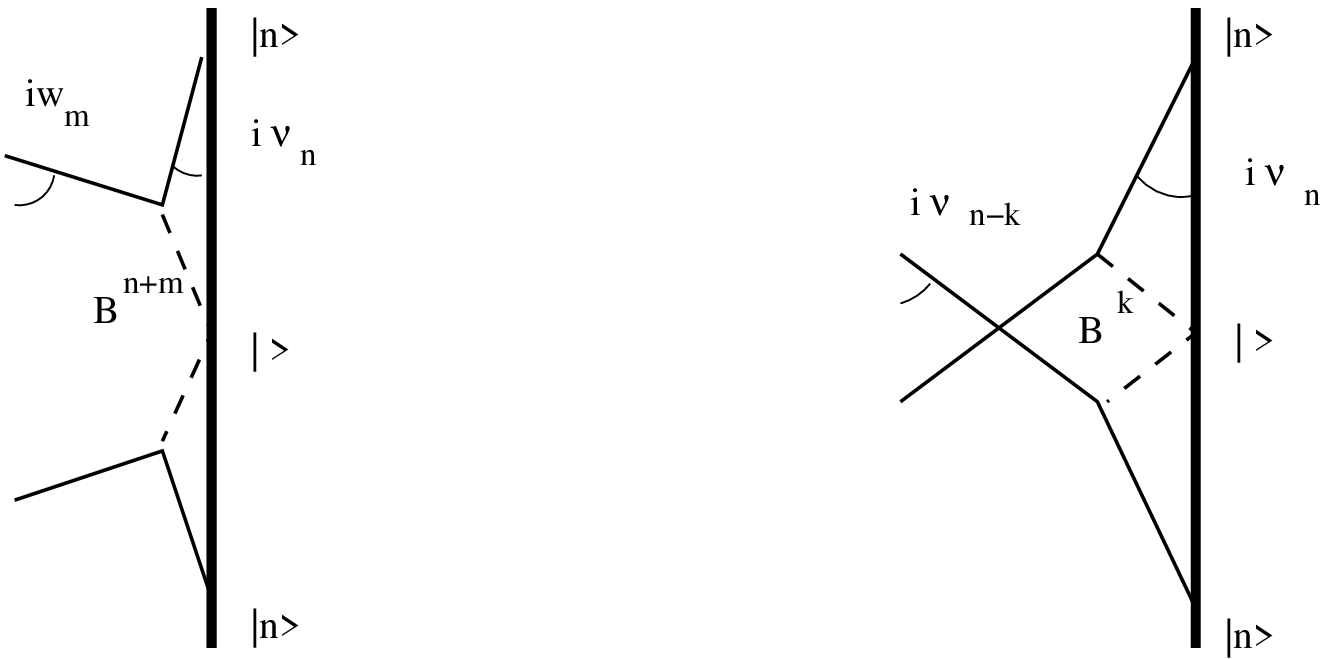}} \par}
\vspace{0.3cm}

\textcolor{black}{The first diagram, however, exists only for \( w_{m}<\nu _{n} \),
since \( B^{n+m} \) must travel towards the boundary. Thus for \( w_{m}>\nu _{n} \)
we have a new boundary boundstate denoted by \( |n,m\rangle : \)
with energy \( m_{|n,m\rangle }=M(\cos (\nu _{n})+\cos (w_{m})) \). }

Repeating these procedures one obtains the following pattern of boundary
excited states and reflection factors: Ground state boundary

\textcolor{black}{\[
|\, \rangle \qquad R(\lambda ,\eta ,\Theta )\quad .\]
The excited states with one index have reflection factors and masses
of the form\[
\begin{array}{c}
|0\rangle \quad \bar{R}(\lambda ,\bar{\eta },\Theta )\\
M\cos (\nu _{0})
\end{array}\dots \begin{array}{c}
|n\rangle \quad \bar{R}(\lambda ,\bar{\eta },\Theta )a_{n}(\eta ,\theta )\\
M\cos (\nu _{n})
\end{array}\quad .\]
The excited boundary states with two indices have reflection factors
and masses as \[
\begin{array}{c}
|n,m\rangle \quad \quad R(\lambda ,\eta ,\Theta )a_{n}(\eta ,\theta )a_{m}(\bar{\eta },\theta )\\
M\cos (\nu _{n})+M\cos (w_{m})
\end{array}\quad .\]
The general state has any of these forms\[
\begin{array}{c}
|n_{1},m_{1},\dots ,n_{k}\rangle \quad \quad \bar{R}(\lambda ,\bar{\eta },\Theta )a_{n_{1}}(\eta ,\theta )a_{m_{1}}(\bar{\eta },\theta )\dots a_{n_{k}}(\eta ,\theta )\\
M\cos (\nu _{n_{1}})+M\cos (w_{m_{1}})+\dots +M\cos (\nu _{n_{k}})
\end{array}\]
\[
\begin{array}{c}
|n_{1},m_{1},\dots ,m_{k}\rangle \quad \quad R(\lambda ,\eta ,\Theta )a_{n_{1}}(\eta ,\theta )a_{m_{1}}(\bar{\eta },\theta )\dots a_{m_{k}}(\bar{\eta },\theta )\\
M\cos (\nu _{n_{1}})+M\cos (w_{m_{1}})+\dots +M\cos (w_{m_{k}})
\end{array}\quad .\]
By finding these boundary excited states the bootstrap is closed in
the sense, that on these boundaries with these reflection factors
every pole can be explained by either a boundary Coleman-Thun diagram
or a boundary bound state creation or both. }

\textcolor{black}{By virtue of their derivation the solution of the
boostrap program contains the parameters of the ground state boundary
reflection factor and has nothing to do with the parameters of the
Lagrangean. Al. B. Zamolodchikov gave the relations of these parameters,
which can be checked both in finite volume analysis and in semiclassical
considerations. \begin{eqnarray*}
\cos \left( \frac{\eta }{\lambda +1}\right) \cosh \left( \frac{\Theta }{\lambda +1}\right)  & = & \frac{M_{0}}{M_{crit}(\beta )}\cos \left( \alpha \right) \quad ;\quad \alpha =\frac{\beta \varphi _{0}}{2}\\
\sin \left( \frac{\eta }{\lambda +1}\right) \sinh \left( \frac{\Theta }{\lambda +1}\right)  & = & \frac{M_{0}}{M_{crit}(\beta )}\sin \left( \alpha \right) \quad ,
\end{eqnarray*}
where, similarly to the bulk model, the actual form of \( M_{crit}(\beta ) \)
depends on the renormalization scheme in which the model is defined.
In the perturbed conformal field theory framework it is \[
M_{crit}(\beta )=m\sqrt{\frac{2}{\beta ^{2}\sin \left( \frac{\beta ^{2}}{8}\right) }}\quad .\]
 }

\section*{Finite volume analysis}

In establishing the relation above one has to compare the exact bootstrap
quantities with computations coming directly form the Lagrangean.
One possible way is to put the system in a finite interval of size
\( L \) and analyze the energy levels of the system as a function
of the system size, imposing two different boundary conditions on
the ends. The two extreme limits can be solved exactly. 

The \( L\to \infty  \) limit is called the infrared (IR) limit. This
is just the theory we solved: The energy eigenstates consists of arbitrary
number of moving solitons, anti-solitons and breathers together with
the boundary excited states corresponding to the two boundaries. 

The \( L\to 0 \) limit is called the ultraviolet (UV) limit. Since
all the potential terms, both the bulk and boundary, are scaled out
the system looks like a free boson\[
H=\frac{1}{8\pi }\int _{0}^{L}\left( \Pi ^{2}+(\partial _{x}\Phi )^{2}\right) dx\]
with compactification radius \( r \) satisfying \( r\beta =\sqrt{4\pi } \).
That is a \( c=1 \) conformal field theory with Neumann boundary
condition applied on both ends. The spectrum can be read off from
the Hilbert space \( a_{-n_{1}}\dots a_{-n_{k}}|n\rangle \, \, ;\, \, \Pi _{0}|n\rangle =\frac{n}{r}|n\rangle  \)
and from the Hamiltonian of the model \[
H=\frac{\pi }{L}\left( 2\Pi _{0}^{2}+\sum _{n\neq 0}na_{-n}a_{n}\right) \quad ;\quad [a_{n},a_{m}]=n\delta _{n+m}.\]
The matching of the IR and UV parameters can be achieved by introducing
a finite volume analysis starting from the UV and another from the
IR with overlaping regions and comparing the energy levels. 

For small \( L \) we regard the boundary sine-Gordon theory as a
joint bulk and boundary perturbation of the boundary conformal field
theory introduced \cite{Neu}. Using the vertex operators of the \( c=1 \)
model\[
V_{n}(x,t)\propto :e^{i\frac{n}{r}\Phi (x,t)}:\quad ;\quad \Psi _{n}(t)=:e^{i\frac{n}{r}\Phi (0,t)}:\]
the perturbation of the Hamiltonian has the form\[
H_{bulk}^{pert.}\to \frac{m^{2}}{2\beta ^{2}}(V_{2}+V_{-2})\quad ;\quad H^{pert.}_{bd.}\to \frac{M_{0}}{2}(e^{-\frac{i}{r}\varphi _{0}}\Psi _{1}+e^{\frac{i}{r}\varphi _{0}}\Psi _{-1})\quad .\]
The computation of the matrix elements of the perturbing potential
is straightforward, but tedious. Truncating the Hilbert space at certain
energy levels and diagonalizing the total Hamiltonian numericaly we
arrive at the Truncated Conform Space Approach (TCSA) which provides
a numerical finite volume spectra being exact for small \( L \). 

For large \( L \) we can obtain a finite volume spectrum by computing
corrections to the IR spectrum. The energy levels of the moving particles
\[
E(L)=\sqrt{M^{2}+P(L)^{2}}\]
 are affected by finite spatial volume. In the case of periodic boundary
condition the momenta are quantized as \[
e^{iPL}=1\, \to P(L)=\frac{2\pi }{L}N\, \, .\]
If, however, we have reflection factors \( R_{0}(P) \) and \( R_{L}(P) \)
on the two ends of the strip, then the momentum quantization for singlet
one particle states changes as \[
e^{i2PL}R_{0}(P)R_{L}(P)=1\to P(L)\quad .\]
As a consequence the finite volume energy levels depend on the reflection
factors and, in the case multiparticle states, also on the scattering
matrices, depending in this way on the IR parameters. 

The comparison between the small \( L \) and large \( L \) regions
either gives a numerical matching between the UV and IR parameters
or if this relation is already conjectured then it provides a numerical
justification of its correctness. We used the second possibility with
the result shown on the following figure. 

\resizebox*{15cm}{!}{\includegraphics{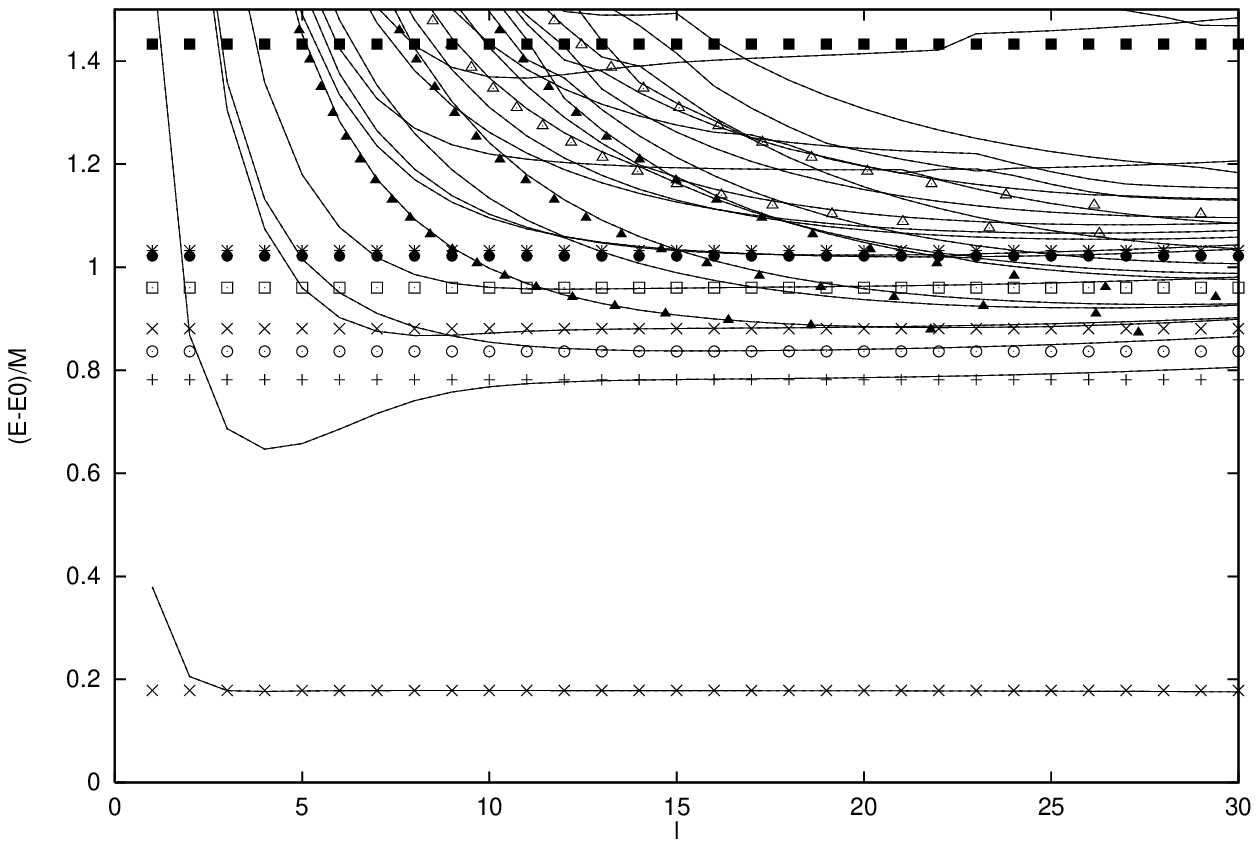}} 

On the figure continous lines come from the TCSA, while the others
correspond to the various multiparticle IR lines. The observed very
good aggrement shows the correctness not only of the UV-IR relation
but also of the entire IR spectrum together with the reflection factors. 

For completeness we mention that in order to derive the exact UV-IR
relation one needs to compute the finite volume energy at least for
one state, say for the ground state, exactly. The thermodynamic Bethe
ansatz provides an integral equation for the ground state energy,
containing the IR parameters, from which the boundary energy can be
extracted \cite{Aup,bsg}. This quantity is related to the vacuum
expectation value of the boundary vertex operator, which can also
be calculated exactly in terms of the UV parameters \cite{FZZ}. The
comparison of the two results gives the required UV-IR relation.

\section*{Semiclassical considerations}

We have seen that the two lowest energy boundary state, the ground
state and the first excited state, characterized by \textcolor{black}{\[
|\, \rangle \quad R(\lambda ,\eta ,\Theta )\quad ;\qquad \begin{array}{c}
|0\rangle \quad \bar{R}(\lambda ,\bar{\eta },\Theta )\\
M\cos (\nu _{0})
\end{array}\]
}are related by the \textcolor{black}{\( s\leftrightarrow \bar{s},\eta \leftrightarrow \bar{\eta } \)
transformations. The classical analogues of these states are} the
two static solutions with lowest energy, given by a static bulk soliton/anti-soliton standing
at the right place' : i.e. by choosing \( \Phi \equiv \Phi _{s}(x,a^{+}) \)
or \( \Phi \equiv \Phi _{\bar{s}}(x,a^{-}) \) for \( x\leq 0 \),
where \[
\Phi _{s}(x,a^{+})=\frac{4}{\beta }{\textrm{arctg}}(e^{m(x-a^{+})}),\qquad \qquad \Phi _{\bar{s}}(x,a^{-})=\frac{2\pi }{\beta }-\Phi _{s}(x,a^{-}),\]

\resizebox*{15cm}{!}{\includegraphics{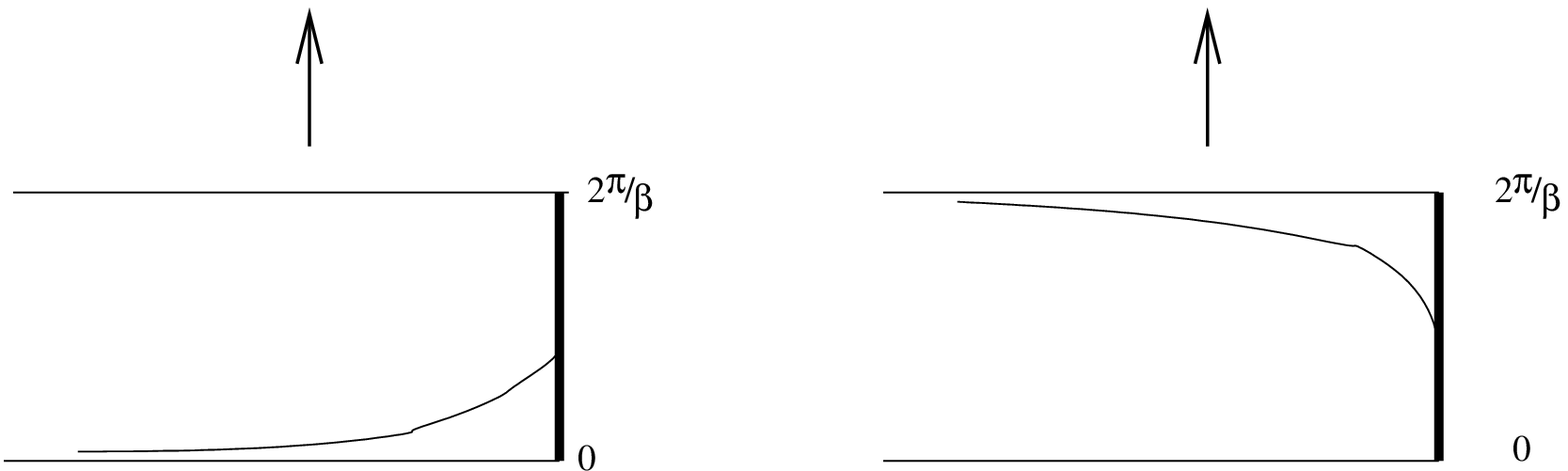}} 

and \( a^{\pm } \) are determined by the boundary condition: \[
\sinh (ma^{\pm })=\frac{A\pm \cos (\alpha )}{\sin (\alpha )}\quad ;\quad A=\frac{4m}{M_{0}\beta ^{2}}\quad .\]
 The energy difference of these two solutions can be written as\begin{equation}
\label{Ediff}
E_{\bar{s}}(M_{0},\varphi _{0})-E_{s}(M_{0},\varphi _{0})=M_{0}(R(+)-R(-))\quad ;\quad R(\pm )=\sqrt{1\pm 2A\cos (\alpha )+A^{2}}\quad .
\end{equation}

In the process of semi-classical quantisation the oscillators associated
to the linearized fluctuations around the static solutions \( \Phi (x,t)=\Phi _{s,\bar{s}}+e^{i\omega t}\xi _{\pm }(x) \)
are quantised. The equations of motion of these fluctuations describe
how the elementary excitations of the field \( \Phi  \) -namely the
first breather- behave in the presence of the nontrivial background.
It can be written as: \begin{equation}
\label{mozg}
\left[ -\frac{d^{2}}{dx^{2}}+m^{2}-\frac{2m^{2}}{\cosh ^{2}(m[x-a^{\pm }])}\right] \xi _{\pm }(x)=\omega ^{2}\xi _{\pm }(x);\qquad x<0\, \, ,
\end{equation}
 where \( \xi _{\pm }(x) \) must satisfy also the linearized version
of the boundary condition: \begin{equation}
\label{lhatfel}
\xi _{\pm }^{\prime }(x)|_{x=0}=-\frac{m}{A}\, \frac{1\pm A\cos \alpha }{R(\pm )}\xi _{\pm }(0)\quad .
\end{equation}
 These eigenvalue problems can be solved exactly by mapping eq.(\ref{mozg})
to a hypergeometric differential equation, whose spectrum in general
has a discrete and a continous part. The discrete real eigenvalues
correspond to excited boundary states, while imaginary eigenvalues
signals the instability of the static solution. The continous spectrum
shows how the first breather reflects on the classical boundary. By
summing up the zero point energies of the quantized fluctuations and
eliminating the logarithmic divergencies by boundary (\( \delta m^{2} \))
and bulk parameter (\( \delta M_{0} \)) renormalization the semiclassically
corrected energy difference can be computed exactly. Performing the
complete analyzis we know semiclassically 

\begin{itemize}
\item the energy of the excited boundary state,
\item the reflection factor of the first breather on the ground state boundary, 
\item and the energy difference between the two lowest lying energy levels. 
\end{itemize}
Comparing these quantities with the semiclassical limit, \( \lambda \to \infty  \),
of their exact quantum values the semiclassical UV-IR parameter correspondence
can be established. If we use the parametrization \[
\eta =\eta _{cl}(\lambda +1)\quad ;\quad \Theta =\Theta _{cl}(\lambda +1)\quad ,\]
then the relation is \begin{equation}
\label{scuvir}
\cos \eta _{cl}=\frac{R(+)-R(-)}{2A}\quad ;\quad \cosh \Theta _{cl}=\frac{R(+)+R(-)}{2A}\quad ,
\end{equation}
which also determines \( M_{crit} \) in the perturbative scheme as
\( M_{crit}/M_{0}=A \). These correspondence can be also be confirmed
by comparing the semiclassical limit of the solitonic reflection factors
with the classical time delay \cite{KP}.

\section*{Boundary resonance states and the stability of the classical solutions}

The stability of a classical solution can be read off from the discrete
spectrum in the semiclassical analyzis. It is convenient to write
\( \omega ^{2}=m^{2}(1-\epsilon ^{2}) \). The normalizable solutions
of eq.(\ref{mozg}) must vanish at \( x\rightarrow -\infty  \), and
assuming \( \epsilon  \) to be positive, they are given by: \[
\xi _{\pm }(x)=Ne^{m\epsilon (x-a^{\pm })}(\epsilon -\tanh [m(x-a^{\pm })])\quad .\]
 The boundary conditions determine the possible values of \( \epsilon  \)
as \[
\epsilon ^{2}+\epsilon \frac{R(\pm )}{A}\pm \frac{\cos \alpha }{A}=0\quad .\]
 It is easy to show, that for the solitonic ground state there is
no positive solution of this equation, while for the anti-solitonic
exited' state one of the roots, namely \begin{equation}
\epsilon =\cos \eta _{cl}\quad ,
\end{equation}
 is positive. In the framework of semi-classical quantisation these
findings imply, that there are no boundary bound states for the ground
state, described by \( \Phi _{s} \), while for the state, described
by \( \Phi _{\bar{s}} \), there is such a boundary bound state. The
semiclassical energy of the boundary state \( \omega _{0}=m\sin \eta _{cl} \)
vanishes for \( \alpha \to 0 \) if \( A>1 \), which shows the instability
of the anti-solitonic boundary solution. This is consistent with the
classical picture, where the energy difference, (\ref{Ediff}), is
precisely the mass of the bulk soliton, and since topological charge
is not conserved in the boundary theory, the higher energy state can
decay into the lower one by emitting a standing soliton.

At this point it is worth comparing the stability analysis of this
\( \alpha \rightarrow 0 \) situation and the one when \( \alpha =0 \)
is set from the start, to emphasize the non smooth nature of the limit.
In the latter case the two classical solutions become \( \Phi _{1}\equiv \frac{2\pi }{\beta } \)
and \( \Phi _{2}\equiv 0 \). The equations for the small fluctuations
are \[
\left[ -\frac{d^{2}}{dx^{2}}+m^{2}\right] \xi _{\pm }(x)=\omega ^{2}\xi _{\pm }(x)\quad ;\quad \xi _{\pm }^{\prime }(x)|_{x=0}=\mp \frac{m}{A}\xi _{\pm }(0)\quad .\]
 Repeating the stability analysis reveals that there are no normalizable
bound state solutions for the ground state, \( \Phi _{2} \), while
for the 'excited' state, \( \Phi _{1} \), there is a normalizable
solution \( Ne^{\frac{m}{A}x} \), with \( \omega ^{2}=m^{2}(1-A^{-2}) \).
When \( A>1 \) this solution signals the existence of a boundary
state, while for \( A<1 \), when this \( \omega ^{2} \) becomes
negative, it indicates the instability of \( \Phi _{1} \). The absolute
value of the purely imaginary frequency is interpreted as the semiclassical
resonance width: \begin{equation}
\label{Gamma}
\Gamma _{cl}=m\sqrt{\frac{1}{A^{2}}-1}\quad .
\end{equation}
 Similarly to the \( \alpha \to 0 \) case analyzed above, we also
have a nice classical interpretation. For this range of \( A \) there
is a moving anti-solitonic solution of the equation of motions\[
\Phi _{s}(x,v)=\frac{4}{\beta }{\textrm{arctg}}\left[ e^{m\left( \frac{x-vt}{\sqrt{1-v^{2}}}-a\right) }\right] \quad ,\quad v=\sqrt{1-A^{2}}\quad ,\]
which looks as follows \vspace{1cm}

\resizebox*{15cm}{!}{\includegraphics{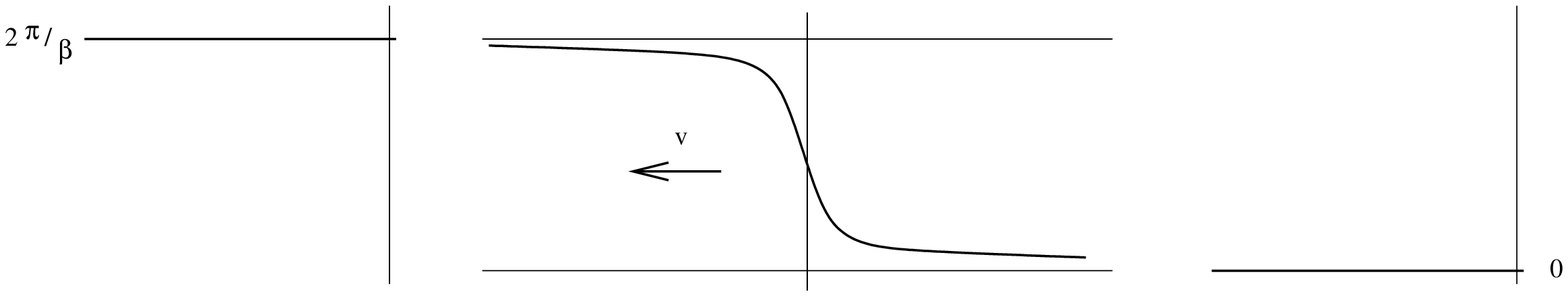}} 

\vspace{0.5cm}

This solution for \( t\to -\infty  \) looks like the excited boundary
state without any anti-soliton, (the anti-soliton is on the nonphysical
part of the space time). For \( t\to \infty  \) the situation changes
as follows: the boundary is in the ground state while an anti-soliton
is moving far away from the boundary. So the excited boundary decayed
into the ground state boundary by emitting a moving anti-soliton.

Let us focus on the quantum theory now. In a theory with bulk a resonance
state the scattering matrix of the stable particles exhibits a pole
singularity at \( s=(M_{res}+i\Gamma /2)^{2} \), where \( M_{res} \)
is the mass, while \( \Gamma  \) the decay width of the resonance.
This can be seen from the form of the bulk propagator \( G(p)^{-1}\propto p^{2}-m^{2} \).
The boundary propagator has the form \( G_{B}(p_{0})^{-1}\propto p_{0}-m \),
thus a boundary resonance state shows up in the reflection factor
as a pole in the energy at \( p_{0}=M_{res}+i\Gamma /2 \), where
\( M_{res} \) is the energy, while \( \Gamma  \) is the decay width
of the resonance state. In order to find the boundary resonance state
we analize the solitonic reflection factors. 

The semiclassical region, where we see the instable state, corresponds
to the \( \alpha =0 \) and \( A>1 \) domain, which can be parametrized
by \( \eta _{cl}=0 \) and \( \Theta _{cl}=\frac{\Theta }{\lambda +1} \)
and can be reached as \( \Theta \to \infty  \), \( \lambda \to \infty  \).
Concentrating on this asymptotic region the \( \frac{\sigma (i\Theta ,\theta )}{\cosh \Theta } \)
term of the reflection factor has simple poles at \[
\theta _{n}=\frac{\Theta }{\lambda }-i\frac{(2n+1)\pi }{2\lambda }\quad ;\quad n\geq 0\quad ,\]
from which the closest to the real axis is \[
\theta _{0}=\frac{\Theta }{\lambda }-i\frac{\pi }{2\lambda }\quad .\]
The energy of this resonance state has a real and an imaginary part
\[
E-E_{0}=M\cosh \left( \frac{\Theta }{\lambda }\right) \cos \left( \frac{\pi }{2\lambda }\right) -iM\sinh \left( \frac{\Theta }{\lambda }\right) \sin \left( \frac{\pi }{2\lambda }\right) \quad ,\]
 which in the semiclassical limit can be written as\[
E-E_{0}=M\cosh \Theta _{cl}-iM\frac{\pi }{2\lambda }\sinh \Theta _{cl}\quad .\]
Using the semiclassical UV-IR relation, (\ref{scuvir}), for \( \eta _{cl}=0 \)
we have \( \cosh (\Theta _{cl})=A^{-1} \). Since the semiclassical
soliton mass is \( M=\frac{8m}{\beta ^{2}}(1-\frac{\beta ^{2}}{2\pi }) \)
the leading order of the real part reproduces the energy difference
(\ref{Ediff}), while the imaginary part the semiclassical decay width
(\ref{Gamma}). 

We have tried to analyze the effect of the boundary resonance state
for the finite volume spectra of the model. We investigated the behaviour
of the solitonic reflection factor near the resonance, but we could
not tune the parameters to keep the resonance strong and obtain a
believable  TCSA spectrum in the same time. Thus the resonance was
unobservable. P. Dorey pointed out, however, that a more significant
effect might be obtained by analyzing the ground state energy of the
system for small volumes similarly to the homogenous sine Gordon case
talk by J. L. Miramontes.

\section*{Conclusions}

We reviewed our recent results on the boundary sine-Gordon model.
By closing the boundary bootstrap we determined the spectrum of boundary
excitations together with the corresponding reflection factors. In
order to check the results we rederived Zamolodchikov`s UV-IR relation
and used it in finite volume analyzis to confirm their correctness.
We also performed a semiclassical quantization, were the correspondence
between a semiclassically unstable static solution and the resonance
pole of the solitonic reflection factors was analyzed in detail. 

The main open problems are the calculation of off-shell quantities
(e.g. correlation functions) and exact finite size spectra. While
correlation functions in general present a very hard problem even
in theories without boundaries, in integrable theories significant
progress was made using form factors. It would be interesting to make
further progress in this direction.

It would also be interesting to work out a formalism (an analogue
of the Cutkosky rules of quantum field theory in the bulk) in which
the rules for the boundary Coleman-Thun diagrams can be justified.
Following \cite{BBT} a work is in progress in this direction. 

Now we are able to report that, as an important step in the supersymmetric
generalisation of the model, the boundary spectrum and reflection
factors have been determined \cite{SSG} by closing the boundary bootstrap.
In order to confirm these results we are developing a TCSA analyzis
for the supersymmetric theory.

\subsection*{Acknowledgements}

We thank M. Kormos and G. Zs. T\'oth for the collaboration in \cite{KP}
and \cite{Genspec}, respectively. Z. B. thanks for the kind hospitality
in Sao Paulo and the useful discussions with Marco Moriconi, Gustav
Delius and Nial MacKay. Z.B. also thanks for the Hungarian National
Science Fund (OTKA) for supporting the travel expences and for the
Bolyai Fellowship. This research was supported in part by the Hungarian
Ministry of Education under FKFP 0043/2001 and by the Hungarian National
Science Fund (OTKA) T037674/02, T034299/01.

\selectlanguage{english}
\end{document}